\documentclass[epj]{svjour}
\usepackage{times}
\usepackage{graphics}
\usepackage{amsmath}
\usepackage{amsfonts}
\usepackage{amssymb}
\usepackage{dsfont}
\usepackage{graphicx}
\usepackage{slashed}
\usepackage{yfonts}
\usepackage{nicefrac}
\usepackage{multirow}
\usepackage[usenames]{color}
\usepackage[colorlinks=true,linkcolor=blue,citecolor=blue,urlcolor=blue]{
hyperref}
\usepackage{cite}
\usepackage{dsfont}

\newcommand{\Slash}[1]{\ooalign{\hfil/\hfil\crcr$#1$}}
\newcommand{\Eq}[1]{(\ref{#1})}

\definecolor{violet}{RGB}{111,0,255}
\definecolor{lred}{rgb}{1,0.90,0.7}
\begin{document}
\title{Hyperon elastic electromagnetic form factors in the 
space-like momentum region}

\author{H\`elios Sanchis-Alepuz\inst{1,2} and Christian S. Fischer\inst{1}   
}                     
\institute{Institut f\"ur Theoretische Physik, Justus-Liebig--Universit\"at 
Giessen, 35392 Giessen, Germany \\ \and
Institut f\"ur Physik, Karl-Franzens--Universit\"at 
Graz, 8010 Graz, Austria}
\date{Received: date / Revised version: date}
%
\abstract{We present a calculation of the electric and magnetic form factors
of ground-state octet and decuplet baryons including strange quarks. We work
with a combination of Dyson-Schwinger equations for the quark propagator and 
covariant Bethe-Salpeter equations describing baryons as bound states of three
(non-perturbative) quarks. Our form factors for the octet baryons are in good
agreement with corresponding lattice data at finite $Q^2$; deviations in 
some isospin channels for the magnetic moments can be explained by missing 
meson cloud effects. At larger $Q^2$ our quark core calculation has predictive 
power for both, the octet and decuplet baryons. 
} 
\maketitle
\section{Introduction}
\label{sec:introduction}
The composite structure of hadronic bound states is encoded in their form factors. 
While there is abundant experimental information on the nucleon electromagnetic structure \cite{Burkert:2004sk,Pascalutsa:2006up}, 
our experimental knowledge of the structure of other baryons and, 
in particular, of those with strange-quark content (hyperons) is scarce and limited to 
some values for static properties \cite{Agashe:2014kda,GoughEschrich:2001ji,Taylor:2005zw,Keller:2011aw,Keller:2011nt}. 
First measurements of hyperon form factors at large 
time-like photon momentum have been presented recently by 
the CLEO collaboration \cite{Dobbs:2014ifa}. The study of the hyperon structure
is also one of the main goals of the CLAS collaboration.

On the theoretical side, there have been many attempts to shed light on the issue of the electromagnetic structure of octet and decuplet 
baryons. Model calculations using bag models \cite{Umino:1991dk}, quark models 
\cite{Darewych:1983yw,Kaxiras:1985zv,Sahoo:1995kg,Sharma:2010vv,Wagner:1995va,
Wagner:1998bu,Wagner:1998fi,Wagner:2000ii,Buchmann:2002xq,VanCauteren:2003hn,
VanCauteren:2005sm,VanCauteren:2005vr,Metsch:2008zz},
instanton models \cite{Merten:2002nz}, QCD sum rules \cite{Liu:2009mb,Aliev:2013jda,Aliev:2013jta,Aliev:2013mda}, Skyrme models \cite{Scoccola:1999ke},
chiral perturbation theory \cite{Kubis:2000aa,Arndt:2003vd,Wang:2008vb} and covariant spectator models \cite{Ramalho:2009vc,Ramalho:2010xj,Ramalho:2011pp,Ramalho:2012ad,Ramalho:2012im,Ramalho:2012pu,Ramalho:2013iaa,Ramalho:2013uza}, show a varying degree of success.
A coherent microscopic description of strong and electromagnetic baryon form factors has
therefore not yet emerged.
Lattice QCD \cite{Leinweber:1990dv,Leinweber:1992hy,Leinweber:1992pv,Boinepalli:2006xd,Lin:2007gv,Boinepalli:2009sq,Shanahan:2014uka,Shanahan:2014cga} is progressively overcoming its limitations and starting to provide more precise \textit{ab initio} data.
While for the octet hyperons these already complement experimental knowledge, corresponding results
for decuplet hyperons are only available for some static observables and are still plagued with 
huge uncertainties. 

In this work we use the combination of Dyson-Schwinger (DSE) and covariant Bethe-Salpeter (BSE) equations to study the elastic electromagnetic form factors of ground-state spin-$\nicefrac{1}{2}$ and  spin-$\nicefrac{3}{2}$ hyperons. The combined framework of DSEs and BSEs provides a rigorously defined connection between QCD as a quantum field theory and hadronic observables. The dynamics of quarks and gluons as constituents of bound states, as well as the interaction vertices among them, is determined by the infinite set of coupled DSEs; these are the necessary elements to define the relevant BSEs, from which solutions one obtains all the properties of hadrons as bound states of quarks and gluons. In practice, for most applications 
the infinite tower of DSEs must be truncated into a finite number of equations. The truncation on the DSEs is non-trivially related to the truncation on the BSEs, such that all the relevant symmetries are preserved; see, e.g. \cite{Sanchis-Alepuz:2015tha} for a review of the formalism. 

The leading-order truncation of the DSE/BSE system is the so-called 
rainbow-ladder (RL) truncation, which is able to reproduce fairly well many 
hadron observables (a selection of results can be found 
in \cite{Eichmann:2013afa} and references therein). In particular, a series of 
recent works 
\cite{Eichmann:2009qa,Eichmann:2011vu,Eichmann:2011pv,SanchisAlepuz:2011jn, 
Sanchis-Alepuz:2013iia,Alkofer:2014bya,Sanchis-Alepuz:2014sca, 
Sanchis-Alepuz:2014wea} have focused on providing a description of ground-state 
baryons as three-quark objects in the RL truncation. This work adds to this 
effort by completing the study of space-like elastic electromagnetic properties 
of the baryon octet and decuplet ground states.

This paper is organised as follows. In section~\ref{sec:formalism} we review the basic elements 
of the DSE/BSE formalism. In section \ref{sec:results} we discuss the assets and limitations of 
the RL scheme and present and discuss results for the electromagnetic properties of 
octet and decuplet hyperons. We summarise and conclude in section \ref{sec:summary}. 
In several appendices we collect technical details of the framework.
 

\section{Review of the formalism}\label{sec:formalism} 

\begin{figure*}[hbtp]
 \begin{center}
  \includegraphics[width=0.8\textwidth,clip]{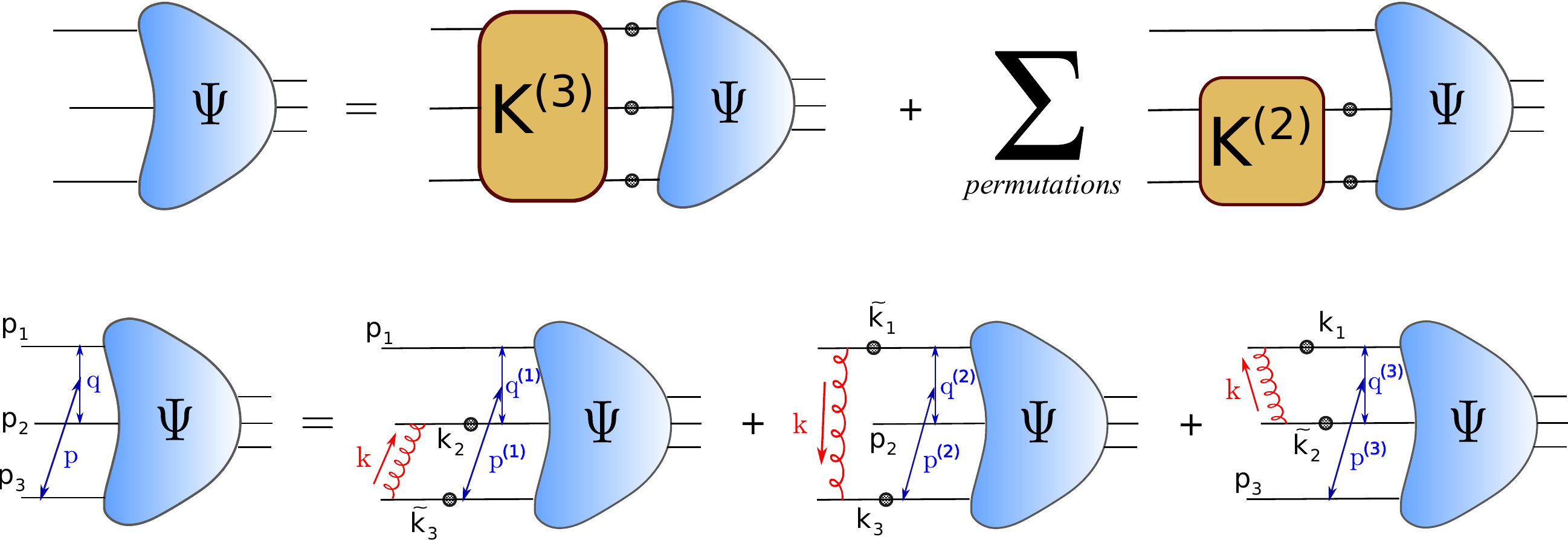}
 \end{center}
 \caption{Upper figure: Diagrammatic representation of the three-body Bethe-Salpeter 
equation with amplitude $\Psi$. Full quark propagators are denoted by straight lines with black dots.
The interaction between the quarks are encoded in the three-body and two-body
kernels $K^{(3)}$ and $K^{(2)}$.
Lower figure: Faddeev equation in the rainbow-ladder 
truncation; see Appendix~\ref{sec:kinematics} for the definitions of  
momenta.}\label{fig:faddeev_eq}
\end{figure*}

In the Bethe-Salpeter framework, a baryon is described by the three-body 
Bethe-Salpeter 
amplitude $\Gamma_{ABCD}(p_1,p_2,p_3)$, where we use $\{ABC\}$ as generic 
indices for spin, flavour and colour indices (e.g. $A\rightarrow\{\alpha,a,r\}$, 
respectively) for the valence quarks and similarly we use $D$ as a collective 
index for the resulting baryon. This amplitude is the solution of the 
three-body Bethe-Salpeter equation Fig.~\ref{fig:faddeev_eq}.
The amplitude depends on the three quark momenta $p_{1,2,3}$, 
which can be 
expressed in terms of two relative momenta $p$ and $q$ and the total momentum 
$P$ (see Eq.~\Eq{eq:defpq} in Appendix~\ref{sec:kinematics}). It is 
decomposed in a tensor product of a spin-momentum part to be 
determined and flavour and colour parts which are fixed
  \begin{equation}\label{eq:BSE_amplitude}
 \Gamma_{ABCD}(p,q,P)=\left(\sum_\rho 
\Psi^\rho_{\alpha\beta\gamma\mathcal{I}}(p,q,P) \otimes 
F^\rho_{abcd}\right)\otimes \frac{\epsilon_{rst}}{\sqrt{6}}~.
\end{equation}
The colour term $\epsilon_{rst}/\sqrt{6}$ fixes the baryon to be a colour singlet 
and the flavour terms $F^\rho_{abcd}$ are the quark-model $SU(3)$-symmetric and mixed
representations (see \cite{Sanchis-Alepuz:2014sca}). The index $\rho$ 
denotes the 
representation of the $SU(3)$ group to which the baryon belongs 
(mixed-symmetric 
or mixed-antisymmetric representation for the baryon octet and symmetric representation 
for the baryon decuplet).

The spin-momentum part of the Bethe-Salpeter amplitude, 
$\Psi^\rho_{\alpha\beta\gamma\mathcal{I}}(p,q,P)$, is a tensor with three Dirac 
indices $\alpha,\beta,\gamma$ associated to the valence quarks and a generic 
index $\mathcal{I}$ whose nature depend on the spin of the resulting bound 
state. They can in turn be expanded in a covariant basis $\{\tau(p,q;P)\}$
\begin{equation}\label{eq:basis_expansion}
\Psi^\rho_{\alpha\beta\gamma\mathcal{I}}(p,q,P)=f^{\rho}_i(p^2,q^2,z_0,z_1,
z_2)\tau_{\alpha\beta\gamma\mathcal{I}}^i(p,q;P)~,
\end{equation}
where the scalar coefficients $\{f\}$ depend on Lorentz scalars $p^2$, $q^2$, 
$z_0=\widehat{p_T}\cdot\widehat{q_T}$, $z_1=\widehat{p}\cdot\widehat{P}$ and 
$z_2=\widehat{q}\cdot\widehat{P}$ only. The subscript $T$ denotes transverse 
projection with respect to the total momentum and vectors with hat are 
normalised. A covariant basis can be obtained using symmetry requirements only; 
for positive-parity spin-$\nicefrac{1}{2}$ baryons it contains 64 elements 
\cite{Eichmann:2009en,Eichmann:2009zx} whereas for spin-$\nicefrac{3}{2}$ 
baryons it contains 128 elements \cite{SanchisAlepuz:2011jn}. In this way, 
one only needs to solve for the scalar functions $f$.

To define the three-body BSE, Fig.~\ref{fig:faddeev_eq}, 
one needs to specify the three-particle  and 
two-particle irreducible kernels, $K^{(3)}$ and  $K^{(2)}$, respectively. In the 
Faddeev approximation the three-body irreducible interactions are 
neglected, and we refer to the simplified BSE as the 
Faddeev equation (FE). In addition one needs to know the full quark propagator 
$S$ (omitting now Dirac indices)
for the quark flavours of interest. In general it can be written as
\begin{equation}\label{eq:full_quark}
 S^{-1}(p)=A(p^2)\left(i\Slash{p}+B(p^2)/A(p^2)\right)~,
\end{equation}
with vector and scalar dressing functions $A(p^2)$ and $B(p^2)$. The ratio 
$M(p^2)=B(p^2)/A(p^2)$ is a renormalisation group invariant and describes the
running of the quark mass with momentum. The dressing functions are obtained as solutions of 
the quark DSE 
\begin{equation}\label{eq:quarkDSE}
 S^{-1}(p)=S^{-1}_{0}(p)+Z_{1f}g^2 C_F \!\!\int_q \!\gamma^\mu
D_{\mu\nu}(p-q)\Gamma^\nu(p,q)S(q)\,\,,
\end{equation}
which also contains the full quark-gluon vertex $\Gamma^\nu$ and the full gluon 
propagator $D_{\mu\nu}$; 
$S_{0}$ is the (renormalised) bare propagator with inverse
\begin{equation}\label{eq:bare_prop}
 S_{0}^{-1}(p)=Z_2\left(i\Slash{p}+m\right)\,,
\end{equation}
where $m$ is the bare quark mass 
and $Z_{1f}$ and $Z_2$ are renormalisation constants. The renormalised strong 
coupling is denoted by $g$ and $\int_q = \int \frac{d^4 q}{(2 \pi)^4}$ abbreviates
a four-dimensional integral in momentum space supplemented with a translationally
invariant regularisation scheme. 

Under certain symmetry requirements, the practical solution of the Faddeev 
equation can be greatly simplified by relating the three two-body interaction diagrams 
in Fig.~\ref{fig:faddeev_eq}. 
As shown in \cite{Eichmann:2011vu,Sanchis-Alepuz:2014sca}, taking the flavour part of the 
Faddeev amplitudes as representations of the $SU(3)$ group induces a specific 
transformation rule for the spin-momentum part under interchange of its 
valence-quark indices. In this case the Faddeev equation for the coefficients 
$f$ reduces to
\begin{flalign}\label{eq:Faddeev_coeff}
 f^{\rho}_i(p^2,q^2,z_0,z_1,z_2)=& \nonumber\\        
C\mathcal{F}_1^{\rho\rho';\lambda}~H_1^{ij}&~g^{\rho'',\lambda}_{j}(p'^2,q'^2,
z'_0,z'_1,z'_2)+ \nonumber\\
C\mathcal{F}_2^{\rho\rho';\lambda}~H_2^{ij}&~g^{\rho'',\lambda}_{j}(p''^2,q''^2,
z''_0,z''_1,z''_2)+ \nonumber\\
C\mathcal{F}_3^{\rho\rho';\lambda}~&~g^{\rho',\lambda}_{i}(p^2,q^2,z_0,z_1,z_2)~,
\end{flalign}
with the colour factor $C=-\nicefrac{2}{3}$ and the flavour matrices $\mathcal{F}$, the rotation 
matrices $H$ and other symbols defined in Appendix \ref{sec:kinematics} (see 
also \cite{Sanchis-Alepuz:2014sca}). 
Here, the index $\lambda$ runs over all elements in a given flavour state (e.g. 
$\lambda=1,2$ for the mixed-symmetric flavour wave function of the proton and 
$\lambda=1,3$ for the mixed-antisymmetric one; see Appendix \ref{sec:flavour}). 
With this simplification, one needs to solve for one of the diagrams only.
For example, particularly simple to solve numerically is 
\begin{flalign}\label{eq:Faddeev_coeff3}
g^{\rho,\lambda}_i(p^2,q^2,z_0,z_1,z_2)= &\nonumber\\
\int_k~\textnormal{Tr}\Bigl[~\bar{\tau}^
{i}_{\beta\alpha\mathcal{I}\gamma}(p,q,P)& K_{\alpha\alpha',\beta\beta'}(p,q,
k) \delta_{\gamma\gamma''} \times  \nonumber\\ 
S^{(\lambda_1)}_{\alpha'\alpha''}(k_1) S^{(\lambda_2)
}_{\beta'\beta''}&(k_2)~\tau^{j}_{\alpha''\beta''\gamma''\mathcal{I}}(p_{(3)},q_{
(3)},P)~\Bigr]\times \nonumber\\ 
&f^{\rho}_j(p_{(3)}^2,q_{(3)}^2,z^{(3)}_0,z^{(3)}_1,z^{(3)}_2)~.
\end{flalign}
where now the quark at the position $\ell$ in each term of the flavour wave 
function is denoted by the superindex $\lambda_\ell$ to keep track of the 
different flavours. The internal relative 
momenta $p_{(3)}$, $q_{(3)}$ (and analogously for $z^{(3)}_0$, $z^{(3)}_1$ and 
$z^{(3)}_2$)  are defined in Appendix \ref{sec:kinematics}. The conjugate 
of the covariant basis $\bar{\tau}$ has been defined in 
\cite{Eichmann:2009en,SanchisAlepuz:2011jn} and it is assumed that the basis 
$\{\tau\}$ is orthonormal.

These considerations are valid for rather general interaction kernels.
However, in general the kernel contains all possible two-body irreducible interactions among quarks 
and for any practical implementation of Eq.~\Eq{eq:Faddeev_coeff3} one must 
truncate it. The BSEs kernels can be related to the integration kernel in the 
quark DSE via functional derivatives \cite{Fukuda:1987su,McKay:1989rk,Munczek:1994zz,Sanchis-Alepuz:2015tha}.
Specifically, if the theory is defined via an nPI effective action, a loop expansion of the action
translates into a truncation of the DSE/BSE system. Moreover, if the truncation is defined in this way,
the requirements of chiral symmetry expressed by axial-vector and vector Ward-Takahashi identities 
leading to massless pions in the chiral limit from the pseudo-scalar meson BSE are automatically fulfilled.
In such a scheme, the leading truncation corresponds to the case in which the full gluon propagator and full quark-gluon 
vertex in \Eq{eq:quarkDSE} are truncated to their tree-level part, and the 
corresponding two-body kernel is a single-gluon exchange with a tree-level 
vector-vector coupling to quarks. In order to generate dynamical chiral-symmetry breaking and be able to reproduce hadron
phenomenology, the interaction is infrared-enhanced with an effective coupling $\alpha_{\textrm{eff}}$ which has to be 
modelled (see sect.~\ref{sec:MT}). This is the rainbow-ladder truncation of the DSE/BSE system. Its salient features are 
simplicity, while maintaining the correct momentum running of the quark propagator and the underlying quark-gluon interaction 
in the medium to large momentum region. Efforts to improve upon the rainbow-ladder scheme can be found e.g. in Refs.~\cite{Bender:1996bb,Watson:2004kd,Bhagwat:2004hn,Matevosyan:2006bk,Fischer:2009jm,Chang:2009zb,Heupel:2014ina,Fischer:2007ze,Sanchis-Alepuz:2014wea}. We will discuss the assets and limitations
of the rainbow-ladder approach below in section \ref{sec:results}.

\subsection{Form factor calculation in the BSE approach}\label{sec:gauging}

\begin{figure*}[hbtp]
 \begin{center}
  \includegraphics[width=0.8\textwidth,clip]{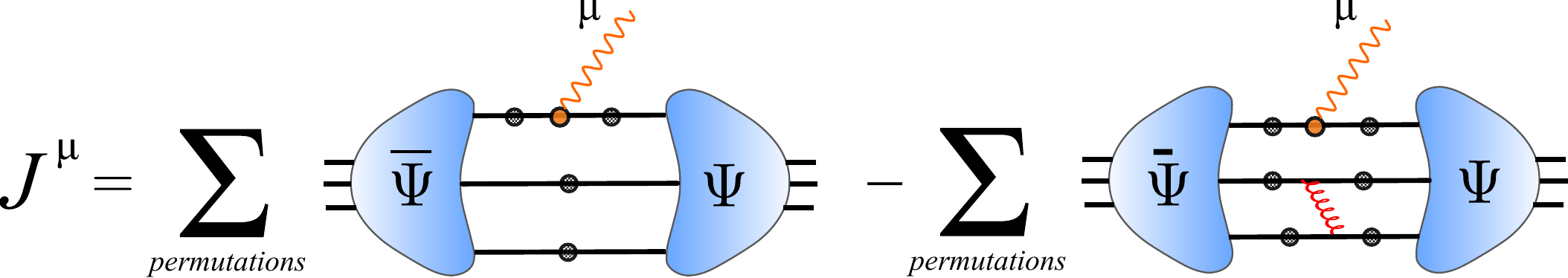}
 \end{center}
 \caption{Diagrams necessary to calculate the current $J^\mu$, describing the coupling of an external photon field to a baryon, 
 described by the Faddeev amplitude $\Psi$, in the RL truncation.}\label{fig:current_diagrams}
\end{figure*}

The procedure to couple an external field to a BSE equation is called \textit{gauging} of the equation and was introduced
in \cite{Haberzettl:1997jg,Kvinikhidze:1998xn,Kvinikhidze:1999xp}. 
The main features of this procedure is that it ensures gauge invariance in the coupling with an external electromagnetic field 
(hence charge conservation) and prevents the over-counting of diagrams.
The specific application of this formalism in the RL-truncated
three-body BSE framework has been already described in \cite{Eichmann:2011vu}. 
We refrain from repeating the steps here and give only
the final expression, corresponding to the diagrams in Fig.~\ref{fig:current_diagrams}, with emphasis in their flavour dependence.

The current $J^\mu$ describing the coupling of a baryon to an external electromagnetic current in the RL truncation is given by
the sum
\begin{flalign}\label{eq:FFeqRL_1} 
J_{\mathcal{I}'\mathcal{I}}^\mu=\sum_{\rho\rho';\lambda}(&\mathcal{Q}^{
\rho\rho';\lambda }_1(J_1^{\rho\rho';\lambda})_{\mathcal{I}'\mathcal{I}}^\mu+	
\mathcal{Q}^{\rho\rho';\lambda}_2(J_2^{\rho\rho';\lambda})_{\mathcal{I}'\mathcal
{I}} ^\mu+\nonumber\\
&\mathcal{Q}^{\rho\rho';\lambda}_3(J_3^{\rho\rho';\lambda})_{\mathcal{I}
'\mathcal{I}} ^\mu) ~,
\end{flalign}
with
\begin{flalign}\label{eq:FFeqRL}
(J_1^{\rho\rho';\lambda})_{\mathcal{I}'\mathcal{I}}^\mu
&=\int_p\int_q\bar{\Psi}^\rho_{\beta'\alpha'\mathcal{I}'\gamma'}(p_f^{\{1\}},q^{\{1\}}_f,P_f)\times\nonumber\\
&\left[\left(S^{(\lambda_1)}(p_1^f)\Gamma^\mu(p_1,Q)S^{(\lambda_1)}(p_1^i)\right)_{\alpha'\alpha}\right.\times\nonumber\\
&\vspace*{0mm}\left.S^{(\lambda_2)}_{\beta'\beta}(p_2)S^{(\lambda_3)}_{\gamma'\gamma}(p_3)\right]\times\nonumber\\
&\vspace*{0mm}\left(\Psi^{\rho'}_{\alpha\beta\gamma\mathcal{I}}(p^{\{1\}}_i,q^{\{1\}}_i,P_i)
                      -\Psi^{\{1\};\lambda}_{\alpha\beta\gamma\mathcal{I}}(p^{\{1 \}}_i,q^{\{1\}}_i,P_i)\right),\nonumber\\
(J_2^{\rho\rho';\lambda})_{\mathcal{I}'\mathcal{I}}^\mu
&=\int_p\int_q\bar{\Psi}^\rho_{\beta'\alpha'\mathcal{I}'\gamma'}(p^{\{2\}}_f,q^{\{2\}}_f,P_f)\left[S^{(\lambda_1)}_{\alpha'\alpha}(p_1)\right.\times\nonumber\\
&\left.\left(S^{(\lambda_2)}(p_2^f)\Gamma^\mu(p_2,Q)S^{(\lambda_2)}(p_2^i)\right)_{\beta'\beta}S^{(\lambda_3)}_{\gamma'\gamma}(p_3)\right]\times\nonumber\\
&\left(\Psi^{\rho'}_{\alpha\beta\gamma\mathcal{I}}(p^{\{2\}}_i,q^{\{2\}}_i,P_i)
                   -\Psi^{\{2\};\lambda}_{\alpha\beta\gamma\mathcal{I}}(p^{\{2 \}}_i,q^{\{2\}}_i,P_i)\right),\nonumber\\
(J_3^{\rho\rho';\lambda})_{\mathcal{I}'\mathcal{I}}^\mu
&=\int_p\int_q\bar{\Psi}^\rho_{\beta'\alpha'\mathcal{I}'\gamma'}(p^{\{3\}}_f,q^{\{3\}}_f,P_f)\times\nonumber\\
&\left[S^{(\lambda_1)}_{\alpha'\alpha}(p_1)S^{(\lambda_2)}_{\beta'\beta}(p_2)\right.\times\nonumber\\
&\left.\left(S^{(\lambda_3)}(p_3^f)\Gamma^\mu(p_3,Q)S^{(\lambda_3)}(p_3^i)\right)_{\gamma'\gamma}\right]\times\nonumber\\
&\left(\Psi^{\rho'}_{\alpha\beta\gamma\mathcal{I}}(p^{\{3\}}_i,q^{\{3\}}_i,P_i)
                   -\Psi^{\{3\};\lambda}_{\alpha\beta\gamma\mathcal{I}}(p^{\{3 \}}_i,q^{\{3\}}_i,P_i)\right),
\end{flalign}
where we have defined\hfill
\begin{align}\label{eq:third_diagram}
 \Psi^{\{1\};\lambda}_{\alpha\beta\gamma\mathcal{I}}=
\int_k  
\widetilde{K}_{\beta\beta'\gamma\gamma'}(k)&S^{(\lambda_2)}_{\beta'\beta''}
(p_2-k)
S^{(\lambda_3)}_{\gamma'\gamma''}(p_3+k) \nonumber\\
&\Psi_{\alpha\beta''\gamma''\mathcal{I}}(p+k,q-k/2,P)~,
\end{align}
as a result of the first term in the Faddeev equation
and similarly we define $\Psi^{\{2\};\lambda}$ and $\Psi^{\{3\};\lambda}$.  The 
transferred photon momentum $Q$ is introduced via
the final and initial momenta of the interacting quark $\kappa$
\begin{equation}
 p_\kappa^{\nicefrac{f}{i}}=p_\kappa\pm\frac{Q}{2}~.
\end{equation}
which also implies $Q=P_f-P_i$. The relative momenta in the respective terms of 
\Eq{eq:FFeqRL} are defined in Appendix~\ref{sec:kinematics}. The charge matrices 
are defined as
\begin{flalign}\label{eq:charge_matrices}
\mathcal{Q}^{\rho\rho';\lambda}_1&=F^\rho_{abc}\textrm{Q}_{aa'}F^{\rho',\lambda}
_{a'bc} \nonumber\\ 
\mathcal{Q}^{\rho\rho';\lambda}_2&=F^\rho_{abc}\textrm{Q}_{bb'}F^{\rho',\lambda}
_{ab'c} \\
\mathcal{Q}^{\rho\rho';\lambda}_3&=F^\rho_{abc}\textrm{Q}_{cc'}F^{\rho',\lambda}
_{abc'} \nonumber
\end{flalign}
with $\textrm{Q}$ the charge operator
\begin{flalign}
 \textrm{Q}=\left(
     \begin{array}{ccc}
      \nicefrac{2}{3} & 0                & 0 \\
            0         & -\nicefrac{1}{3} & 0 \\      
            0         & 0                & -\nicefrac{1}{3}
     \end{array}\right)~.
\end{flalign}

The three terms in \Eq{eq:FFeqRL_1} can be related to each other in the same 
fashion as Eq.~\Eq{eq:Faddeev_coeff}, so that only one term needs to be solved 
explicitly. It can be shown \cite{Eichmann:2011vu} that 
\begin{flalign}\label{eq:FFeqRL_simple}
J_{\mathcal{I}'\mathcal{I}}^\mu=\sum_{\rho\rho';\lambda}(&\mathcal{Q}^{\rho\rho'
}_1(\mathcal{F}_1^{\rho\rho'}J_3^{\rho'\rho'';\lambda}\mathcal{F}_1^{T,
\rho''\rho})_{ \mathcal{I}'\mathcal{I}}^\mu+\nonumber\\
&\mathcal{Q}^{\rho\rho'}_2(\mathcal{F}_2^{\rho\rho'}J_3^{\rho'\rho'';\lambda}
\mathcal{F}_2^{T,\rho''\rho})_{\mathcal{I}'\mathcal{I}}^\mu+\nonumber\\
&\mathcal{Q}^{\rho\rho'}_3(J_3^{\rho\rho';\lambda})_{\mathcal{I}'\mathcal{I}}
^\mu)~,
\end{flalign}
with the matrices $\mathcal{F}$ and the results of the flavour contractions above for the different baryons given in Appendix~\ref{sec:flavour}.

Finally, the quark-photon vertex $\Gamma^\mu$ is calculated from an
inhomogenous Bethe-Salpeter equation
\begin{flalign}
 \Gamma^\mu(p,Q)&=iZ_2\gamma^\mu\nonumber\\                
&+\int_kK_{q\bar{q}}\left(S(k+Q/2)\Gamma^\mu(k,Q)S(k-Q/2)\right)~,
\end{flalign}
and using for $K_{q\bar{q}}$ the RL kernel \Eq{eq:RLkernel} with $C=4/3$ and
for the quark propagator $S$ the solutions of the RL-truncated quark DSE.

\subsection{Effective coupling of one-gluon exchange}\label{sec:MT}

For the effective interaction in the RL truncation we use 
the Maris-Tandy model~\cite{Maris:1997tm,Maris:1999nt} which has been employed 
frequently in hadron studies within the rainbow-ladder BSE/DSE framework. 
The advantages of this model are its simplicity and that it performs very well 
for phenomenological calculations of ground-state meson and baryon properties
in most channels. It can be defined via the combination of the relevant parts 
in the quark self-energy
\begin{align}
Z_{1f} C_F\frac{g^2}{4\pi} D_{\mu\nu}(k)\Gamma^\nu(p,q) = Z_2^2C_FT_{\mu\nu}^k 
\frac{\alpha_{\mathrm{eff}}(k^2)}{k^2}\gamma^\nu\;,
\end{align}
with $\alpha_{\mathrm{eff}}(k^2)$ the effective coupling. In this way 
the resulting kernel in the BSE is given by
\begin{equation}\label{eq:RLkernel}
	{K}^{RL}_{\alpha\alpha'\beta'\beta}(k)= -4\pi C_F~Z_2^2
~\frac{\alpha_{\textrm{eff}}(k^2)}{k^2}~
	T_{\mu\nu}(k)~\gamma^\mu_{\alpha\alpha'}  \gamma^\nu_{\beta'\beta}\,\,.
\end{equation}
with gluon momentum $k=p-q$ and transverse projector 
$T_{\mu\nu}(k)=\delta^{\mu\nu}-\hat{k}^\mu \hat{k}^\nu$.
The effective running coupling $\alpha_{\textrm{eff}}$ is finally given by
\begin{flalign}\label{eq:MTmodel}
\alpha_{\textrm{eff}}(k^2) {}=&
 \pi\eta^7\left(\frac{k^2}{\Lambda^2}\right)^2
e^{-\eta^2\frac{k^2}{\Lambda^2}}\nonumber\\ &+{}\frac{2\pi\gamma_m
\big(1-e^{-k^2/\Lambda_{t}^2}\big)}{\textnormal{ln}[e^2-1+(1+k^2/\Lambda_{QCD}
^2)^2]}\,. 
\end{flalign}
This interaction reproduces the one-loop QCD behaviour of the quark propagator at 
large momenta and the Gaussian distribution of interaction strength in the intermediate 
momentum region provides enough strength for dynamical chiral symmetry breaking to take place. 
The scale $\Lambda_t=1$~GeV is introduced for technical reasons and has no impact on
the results. For the anomalous dimension we use $\gamma_m=12/(11N_C-2N_f)=12/25$,
corresponding to $N_f=4$ flavours and $N_c=3$ colours. The scale in the ultraviolet 
part of the coupling is set to $\Lambda_{QCD}=0.234$ GeV. 
\begin{table*}[t!]
 \begin{center}
 \small
\renewcommand{\arraystretch}{1.2}
\caption{Hyperon masses (in GeV) from the rainbow-ladder truncated Faddeev equation. 
We give the central value of the bands corresponding to a variation of 
$\eta$ between $1.6$ and  $2.0$ with the halfwidth of the bands added in 
brackets. 
We compare also with experimental values \cite{Beringer:1900zz}.
\label{tab:masses}}
\begin{tabular}{c||cccc}\hline
  &  $\Sigma / \Lambda$	&  $\Xi$     & $\Sigma^*$ 	& $\Xi^*$  \\ \hline\hline
 Faddeev &  1.073 (1) 	&  1.235 (5) &  1.33 (2) 	& 1.47(3) \\ \hline
 Experiment     & 1.189/1.116 & 1.315 & 1.385 (2)	& 1.533 (2)		\\ \hline
 Relative difference	& 10/4 \% &  6 \%  &  4 \%		& 4 \% 	\\ 
\hline
\end{tabular}
 \end{center}
\end{table*}

In the infrared momentum region, the interaction strength is characterised by a 
scale $\Lambda$ and a dimensionless parameter $\eta$ that controls the width of 
the interaction. 
The scale $\Lambda=0.72$~GeV is adjusted to reproduce the 
experimental pion decay constant from the truncated pion BSE.
This as well as many other pseudo-scalar ground-state observables, such as the masses of 
ground-state mesons and baryons, turn out to 
be almost insensitive to the value of $\eta$ in the range of values of $\eta$ 
between $1.6$ and $2.0$ see, e.g. 
\cite{Krassnigg:2009zh,Nicmorus:2010mc,Eichmann:2011vu}). 
The $u/d$ and $s$ current-quark masses are 
fixed to reproduce the physical pion and kaon masses, respectively. 
The corresponding values are $m_{u/d}(\mu^2)=3.7$~MeV and 
$m_s(\mu^2)=85$~MeV. The renormalisation scale is chosen to be $\mu^2= (19 
\,\mbox{GeV})^2$.


\section{Results}\label{sec:results}

We have calculated the electromagnetic elastic form factors for all lowest-lying hyperons of the spin-$\nicefrac{1}{2}$ isospin octet and spin-$\nicefrac{3}{2}$ isospin decuplet.
In the plots below we present (coloured bands) the numerical results for values of the $\eta$ parameter between $\eta=1.6$ and  $\eta=2.0$.  Moreover, we show fits of our results using a generalised multipole expression
\begin{equation}
 G(Q^2)=\frac{n_0+n_1Q^2}{1+d_1Q^2+d_2Q^4+d_3Q^6+d_4Q^8}\,.
\end{equation}
For analogous calculations of form factors of non-strange baryons and the 
Omega resonance we refer the reader to 
\cite{Eichmann:2011vu,Sanchis-Alepuz:2013iia}.

\begin{figure}[b!]
\begin{center}
\resizebox{0.45\textwidth}{!}{\includegraphics{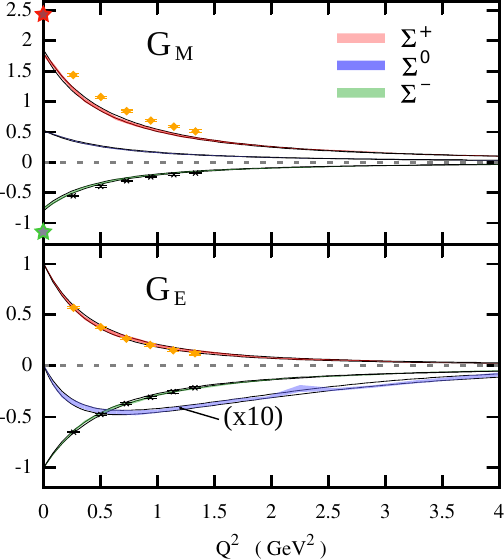}} 
\end{center}
\caption{Magnetic (upper panel) and electric (lower panel) form factors for the 
octet Sigma triplet. 
The magnetic form factor is given in units of the nuclear magneton $\mu_N$.
Coloured bands represent the result of the numerical calculation for $\eta=[1.6,2.0]$.
Stars indicate experimental values. For finite $Q^2$ we also include lattice data from \cite{Shanahan:2014uka,Shanahan:2014cga}.
Solid black lines are our best fits for $\eta=1.6$ and $\eta=2.0$.}
\label{fig:sigma_elastic}       
\end{figure}

Before we discuss the results it is worth summarising the main strengths and weaknesses of the rainbow-ladder truncation scheme used in the 
present calculation, in order to facilitate the subsequent interpretation of results:
\begin{itemize}
\item Ground-state baryon masses obtained from the present formalism in the RL truncation, in combination with the Maris-Tandy model from 
  Eq.~\Eq{eq:MTmodel}, show excellent agreement with experimental data. In the case of baryons with strange quarks the small differences 
  can be attributed to the lack of quark-mass dependence of the interaction (see table \ref{tab:masses}).
\item Results for nucleon, Delta and Omega electromagnetic form factors \cite{Eichmann:2011vu,Sanchis-Alepuz:2013iia} are consistent with 
  a quark-core picture. That is, since meson-cloud effects are not taken into account, form factors for low-$Q^2$ are consistently underestimated.
  In the high-$Q^2$ region, where the effect of the meson cloud is small, the agreement with experimental and/or lattice data is however excellent.
\item With our current computational techniques, calculations are limited to a maximum value of $Q^2$ due to the presence of complex conjugate 
  poles in the quark propagators (see Appendix~\ref{sec:kinematics} and e.g. 
\cite{Eichmann:2011aa,Dorkin:2013rsa}). The specific values of these limits 
depend on the 
masses of the baryons of interest. Nevertheless in most cases
  results are still smooth beyond these regions. This can be seen e.g. in the electric form factor of the $\Sigma_0$, Fig.~\ref{fig:sigma_elastic},
  for $Q^2 \ge 2.25$ GeV$^2$. With due caution, extrapolations for high-$Q^2$ values are thus possible.
\end{itemize}

\subsection{Octet hyperons}

The elastic coupling of a spin$-\nicefrac{1}{2}$ baryon to an external electromagnetic field is described by the following current
\begin{align}\label{eq:D_current}
J^\mu\left(Q,P\right) = i\Lambda_+(P_f) & \left(F_1(Q^2)\gamma^\mu \right.\nonumber\\
                                        & \left.-\frac{F_2(Q^2)}{2M}\sigma^{\mu\nu}Q^\nu\right)\Lambda_+(P_i)\,,
\end{align}
parametrized by the two Dirac form factors $F_1$ and $F_2$ and with 
$\Lambda^+(\hat{P})=\left(\mathds{1}+\Slash{\hat{P}}\right)/2$ 
the positive-energy projector.
We present here results for the electric and magnetic Sachs form factors
\begin{flalign}
 G_E(Q^2)&=F_1(Q^2)-\frac{Q^2}{2 M^2}F_2(Q^2)~,\\
 G_M(Q^2)&=F_1(Q^2)+F_2(Q^2)~.
\end{flalign}
In the case of the baryon octet, experimental data exist for the magnetic moment of all members of the octet 
and for the electric radius of the nucleon and the $\Sigma^-$. In addition, precise lattice QCD data for 
non-vanishing momentum transfer up to $Q^2=1.3~$GeV$^2$ are now available \cite{Shanahan:2014uka,Shanahan:2014cga}.

\begin{figure}[t!]
\begin{center}
\resizebox{0.45\textwidth}{!}{%
  \includegraphics{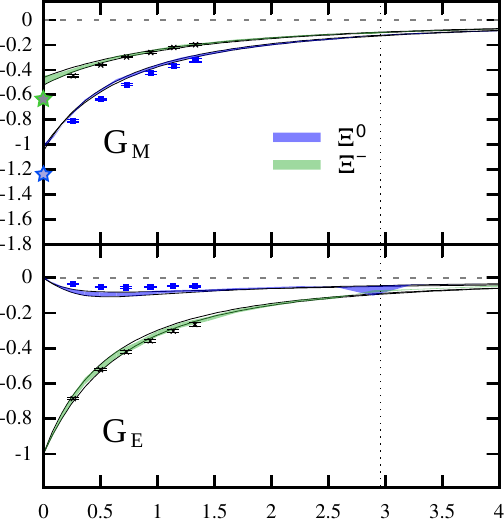}} 
\end{center}
\caption{Magnetic (upper panel) and electric (lower panel) form factors for the 
octet Xi doublet.
The magnetic form factor is given in units of the nuclear magneton $\mu_N$.
Coloured bands represent the result of the numerical calculation for $\eta=[1.6,2.0]$.
Stars indicate experimental values. For finite $Q^2$ we also include lattice data from \cite{Shanahan:2014uka,Shanahan:2014cga}.
Solid black lines are our best fits for $\eta=1.6$ and $\eta=2.0$. 
Vertical dotted lines indicate the region beyond which the singularities of 
the quark propagator are probed.}
\label{fig:xi_elastic}       
\end{figure}

\begin{figure}[t!]
\begin{center}
\resizebox{0.45\textwidth}{!}{%
  \includegraphics{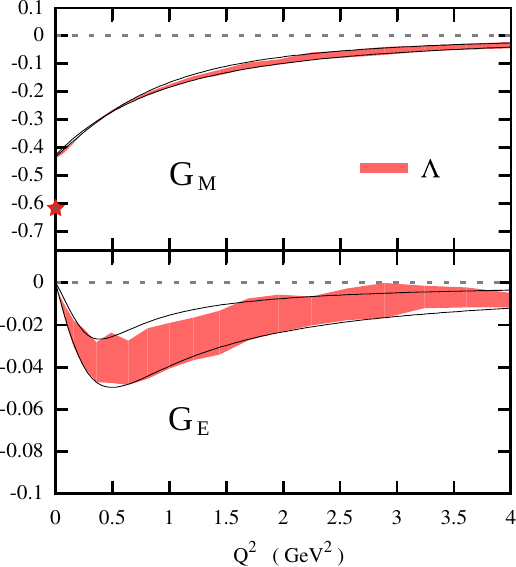}
} 
\end{center}
\caption{Magnetic (upper panel) and electric (lower panel) form factors for the 
Lambda singlet.
The magnetic form factor is given in units of the nuclear magneton $\mu_N$.
Coloured bands represent the result of the numerical calculation for $\eta=[1.6,2.0]$.
Solid black lines are our best fits for $\eta=1.6$ and $\eta=2.0$.}
\label{fig:lambda_elastic}       
\end{figure}

In figures \ref{fig:sigma_elastic}, \ref{fig:xi_elastic} and \ref{fig:lambda_elastic} we show 
the electric and magnetic form factors of the Sigma and Xi multiplets and the Lambda singlet. 
We compare our results with experimental data for the magnetic moments \cite{Agashe:2014kda} 
and with lattice QCD data determined from $N_f=2+1$ dynamical configurations under the omission 
of disconnected diagrams. The lattice results have been obtained at unphysical pion masses and 
extrapolated to the physical point using partially quenched chiral perturbation theory
\cite{Shanahan:2014uka,Shanahan:2014cga}. The systematic error associated with this setup is 
arguably smaller than the one induced by the omission of meson cloud effects in our results.

For the charged $\Sigma$ baryons and for the $\Xi$ doublet our results show an overall good 
agreement with the corresponding lattice data. This is particularly true for the electric 
form factors, which are protected in the infrared by charge conservation. In the magnetic sector
we observe an interesting pattern. While $G_M$ for the $\Sigma^-$ and $\Xi^-$ agree well with 
the lattice data at finite $Q^2$, there is a significant deviation for the $\Sigma^+$ and 
the $\Xi^0$. At $Q^2=0$ the deviation of our magnetic moment with the experimental
value is similar in relative size for the $\Sigma^+$ and the $\Sigma^-$, whereas it is 
much smaller for the $\Xi^+$ than for the $\Xi^0$, see Tab.~\ref{tab:magnetic_moments_octet}.

This pattern may be understood from
the different influence and the interplay of pion and kaon cloud effects on the various 
states \cite{Leinweber:2002qb,Boinepalli:2006xd}. For the $\Sigma^\pm$-states, pion cloud 
effects play an important role for both states at small $Q^2$, leading to the observed
discrepancies of our magnetic moments with the experimental values. For larger values of 
$Q^2$ also kaon cloud effects are important, which are strong only in the $\Sigma^+$-channel.
This may explain the better agreement of our results with the lattice data for $\Sigma^-$.
For the $\Xi$-states, pion cloud effects are small and kaon cloud effects are much larger 
in the $\Xi^0$ than in the $\Xi^-$. As a consequence we observe only small deviations of 
the magnetic moment of the $\Xi^-$ with experiment and also much better agreement with the
lattice data than for the $\Xi^0$. 

\begin{table}[b!]
\caption{Magnetic moments $\mu$ (in units of the nuclear magneton $\mu_N$) of the octet hyperons, 
compared to experimental values \cite{Agashe:2014kda}. Electric and magnetic mean-square radii 
(in fm$^2$) of the octet hyperons.}
\label{tab:magnetic_moments_octet}       
\begin{center}
\begin{tabular}{c|c|c||c|c}
			& 	$\mu$		&	$\mu_{exp}$ 	&	$\langle r^2_E \rangle$	&	$\langle r^2_M \rangle$	\\ [1.0ex] \hline 
$\Sigma^+$	& 	1.82(2) 	&   2.46(1) 		& 0.56(3)					& 0.43(2)					\\ [1.0ex] 
$\Sigma^0$	& 	0.521(1) 	&	 --   			& 0.057(8)					& 0.39(3)					\\ [1.0ex]   
$\Sigma^-$	& 	-0.78(2) 	&	-1.16(3)  		& 0.45(3)					& 0.50(1)					\\ [1.0ex]   
$\Xi^0$ 	&	-1.05(1) 	&	-1.250(14) 		& 0.10(1)					& 0.35(3)					\\ [1.0ex]  
$\Xi^-$ 	&	-0.57(4) 	&	-0.651(3) 		& 0.37(4)					& 0.20(2)					\\ [1.0ex]  
$\Lambda$ 	&	-0.435(5)	&	-0.613(4)		& 0.04(1)					& 0.21(1)
\end{tabular}
\end{center}
\end{table}

Clearly, in the future we need to include the effects of the pion and kaon cloud in our 
calculations. First steps in this direction have been performed on the level of meson and 
baryon masses in Refs.~\cite{Fischer:2008sp,Fischer:2008wy,Sanchis-Alepuz:2014wea}. The
generalisation of this framework to include form factors is under way.

For high-$Q^2$ values the effect of the meson cloud diminishes and we can see that in Figs.~\ref{fig:sigma_elastic} 
and \ref{fig:xi_elastic} the agreement with lattice data improves. This fact encourages us to consider our results 
as a good representation of hyperon form factors for $Q^2$ values which are inaccessible to lattice QCD with the 
present computing capabilities. It, moreover, supports the predictive value of our results, for moderate to 
high-$Q^2$, in those cases in which lattice data is not available, such as for $\Sigma^0$ and $\Lambda$ baryons 
or for decuplet hyperons (see below).

It is straightforward in our framework to extract the corresponding electric or magnetic distribution radius using the formula
\begin{equation}\label{eq:radius}
\langle r^2\rangle_G=-\frac{6}{G(0)}\left.\frac{dG(Q^2)}{dQ^2}\right|_{Q^2=0}~,
\end{equation}
where the term $G(0)$ is set to one for form factors constrained to vanish at the origin. The results for the electric and 
magnetic square radii are shown in Tab.~\ref{tab:magnetic_moments_octet}. For these observables, only the electric radius 
of $\Sigma^-$ is known experimentally, $r^2_{E,\Sigma^-}=0.60~$fm$^2$. As expected with the absence of a long range meson 
cloud, our results underestimate the experimental value. Even though a quantitative prediction of charge and magnetic radii 
will have to wait until meson cloud effects can be taken into account in our approach, there are a number of qualitative 
features worth discussing, since these will presumably survive in a more complete calculation:
Despite having equal quark content, the electric radius of $\Sigma^0$ is larger than that of the $\Lambda$. In constituent 
quark models this is usually interpreted in a two-step way: in the $\Lambda$ a scalar, isoscalar diquark is formed between 
the non-strange quarks whereas in the $\Sigma^0$ the scalar diquark is formed between strange and non-strange quarks. 
In the latter case, repulsive hyperfine interaction in the non-strange sector makes the charge distribution broader. As 
already discussed in \cite{Boinepalli:2006xd}, confirmed in \cite{Boinepalli:2009sq} and also in our calculation 
(see Tab.~\ref{tab:radii_decuplet} below), this explanation is in contradiction with the fact that decuplet hyperons are 
slightly smaller than their octet counterparts. In the decuplet, there is no reason why scalar diquarks should dominate 
and thus charge distributions should be broader, which is not the case. Thus one needs to resort to alternative explanations.
In \cite{Boinepalli:2006xd} the difference between the electric radii of $\Sigma^0$ and $\Lambda$ is associated to the 
different intermediate virtual transitions influencing each baryon. We do not include such intermediate states in 
our present calculation, but still see this difference. Thus from our framework we conclude that the difference may be 
merely due to the different flavour symmetry of both states.

It is also interesting to note how for the magnetic radii some patterns are reversed. Whereas the electric distribution 
is broader in the $\Sigma^+$ than in the $\Sigma^-$ and also in the $\Xi^0$ than in $\Xi^-$, the opposite occurs for the 
magnetic distribution. This pattern is also observed in lattice calculations \cite{Boinepalli:2006xd,Shanahan:2014uka}. 
Moreover, $\Sigma^-$ has the largest magnetic radius of all octet members, a feature also seen in \cite{Boinepalli:2006xd,Shanahan:2014uka}.

Finally we wish to mention that in contrast to all other form factors, it was impossible to fit the $\Sigma^0$ form 
factors with a dipole-like expression. In order to obtain a good representation we had to include terms up to 
${\sim}Q^8$ in the denominator of Eq.~\Eq{eq:fits}, see Tab. \ref{tab:fit_octet_elastic} in the appendix.

\subsection{Decuplet hyperons}

The elastic coupling of a spin$-\nicefrac{3}{2}$ baryon to an external 
electromagnetic field is described by the following current
\begin{align}\label{eq:RS_current}
 J^{\mu,\alpha\beta}(P,Q)={}&\mathbb{P}^{\alpha\alpha'}(\hat{P}_f)\left[
\left((F_1+F_2)i\gamma^\mu-F_2\frac{P^\mu}{M}\right)\delta^{\alpha'\beta'}
\right.\nonumber \\                         
&+\left.\left((F_3+F_4)i\gamma^\mu-F_4\frac{P^\mu}{M}\right)\frac{Q^{\alpha'}Q^{
\beta'}}{4M^2}\right]
\nonumber\\
& ~ \hspace{39mm} \mathbb{P}^{\beta'\beta}(\hat{P}_i)
\end{align}
where $\mathbb{P}$ is the Rarita-Schwinger projector 
\begin{align}\label{eq:def_projectors}
 \mathbb{P}_+^{\mu\nu}(\hat{P})=&~\Lambda_+(\hat{P})
 \left(T_P^{\mu\nu}-\frac{1}{3}\gamma^\mu_T\gamma^\nu_T\right)~.
\end{align}
The current is now parameterised by the four form factors $F_i$, $i \in\{1,...,4\}$.
We present results in terms of the Sachs form factors known as the electric monopole
($G_{E_0}(Q^2)$), magnetic dipole ($G_{M_1}(Q^2)$), electric quadrupole
($G_{E_2}(Q^2)$) and magnetic octupole ($G_{M_3}(Q^2)$) form factors, defined by
\begin{align}
G_{E_0} &= \left(1+\frac{2\tau}{3}\right) ( F_1 - \tau F_2)  - \frac{\tau}{3}
(1+\tau) \,( F_3 - \tau F_4) \,, \\
G_{M_1} &= \left(1+\frac{4\tau}{5}\right) (F_1+F_2) - \frac{2\tau}{5} (1+\tau)\,
(F_3 + F_4)\,,\\
G_{E_2} &= (F_1 - \tau F_2) - \frac{1}{2}\,(1+\tau) \,(F_3 - \tau F_4)\,,\\
G_{M_3} &= (F_1 + F_2) - \frac{1}{2}\,(1+\tau) \,(F_3 + F_4)\,,
\end{align} 
with $\tau=Q^2/4M^2$.

Whereas for the Delta baryon lattice QCD data of the $Q^2$ evolution of form factors exists, 
for the strange members of the decuplet only static properties have been calculated on the lattice \cite{Boinepalli:2009sq}. 
Considering the good comparison with lattice data in the case of octet baryons in this paper 
and in Ref.~\cite{Eichmann:2011vu} as well as for the Delta in Ref.~\cite{Sanchis-Alepuz:2013iia}, 
we can regard the results presented below as qualitative predictions for moderate $Q^2$ which become 
quantitative predictions at higher $Q^2$.

\begin{figure}
\begin{center}
\resizebox{0.46\textwidth}{!}{%
  \includegraphics{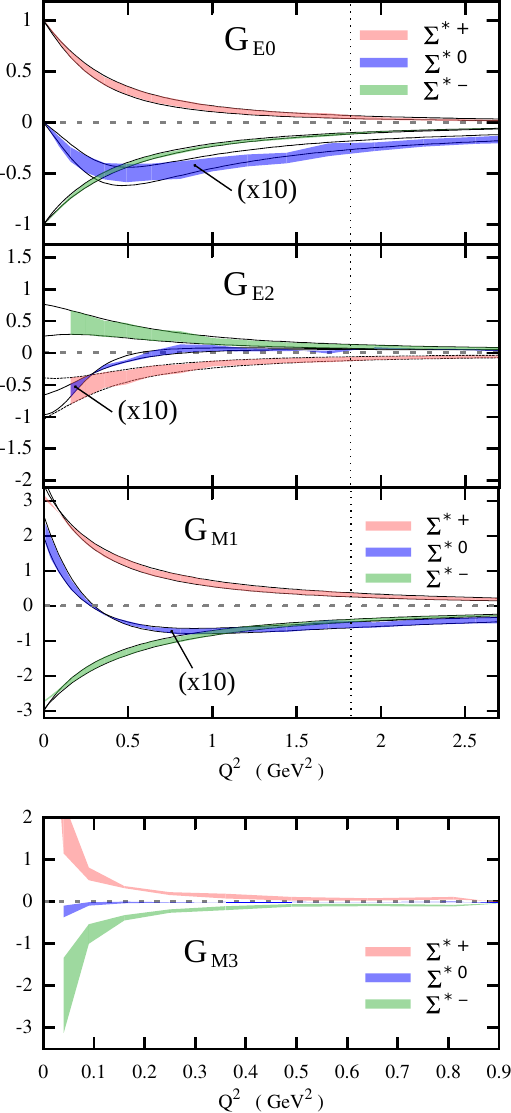}
} 
\end{center}
\caption{Electric monopole, electric quadrupole, magnetic dipole and magnetic octupole 
form factors for the Sigma triplet.
Coloured bands represent the result of the numerical calculation for $\eta=[1.6,2.0]$.
Solid black lines are our best fits for $\eta=1.6$ and $\eta=2.0$.
Vertical dotted lines indicate the region beyond which the singularities of 
the quark propagator are probed.}
\label{fig:sigma32_GE}       
\end{figure}

\begin{figure}
\begin{center}
\resizebox{0.45\textwidth}{!}{%
  \includegraphics{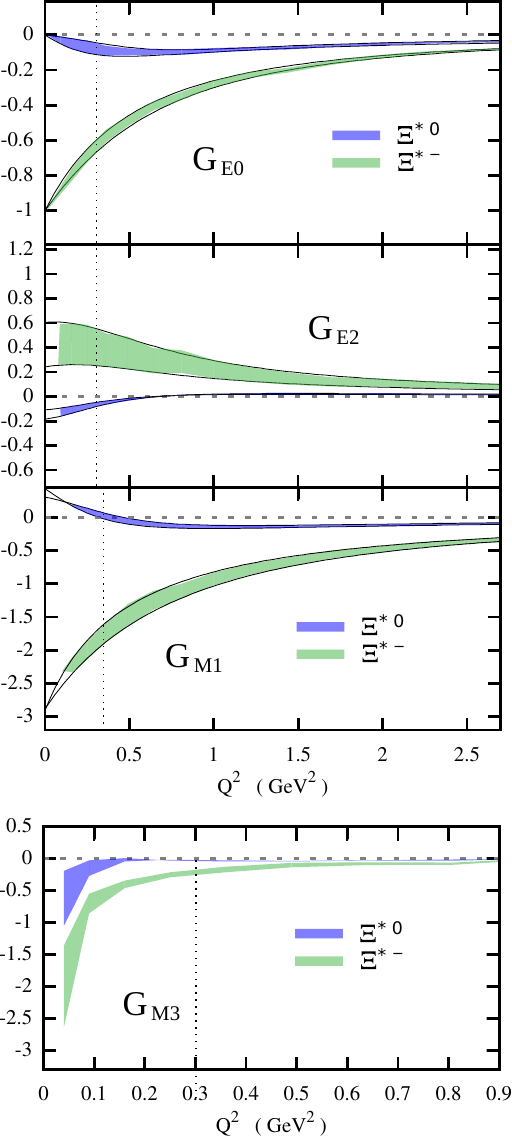}
} 
\end{center}
\caption{Electric monopole, electric quadrupole, magnetic dipole and magnetic octupole
form factors for the Xi doublet. 
Coloured bands represent the result of the numerical calculation for $\eta=[1.6,2.0]$.
Solid black lines are our best fits for $\eta=1.6$ and $\eta=2.0$.
Vertical dotted lines indicate the region beyond which the singularities of 
the quark propagator are probed.}
\label{fig:xi32_GE}       
\end{figure}

In Fig.~\ref{fig:sigma32_GE} we show our results for the electric monopole $G_{E_0}$ and 
quadrupole $G_{E_2}$ form factors. It is interesting to note that the form factors 
corresponding to $\Sigma^{*~0}$ are not vanishing. Moreover, the electric monopole form 
factor is of the same order as the analogous $G_E$ of its octet counterpart. 
This is 
in contrast with the Delta baryon multiplet. In that case, $SU(2)$ isospin symmetry implies 
(in the case of a flavour diagonal interaction like the RL kernel) that the form factors 
are proportional to the baryon charge, thus implying that all form factors of the $\Delta^0$ 
vanish identically.

The electric quadrupole form factor provides a measure of the deviation of the 
charge distribution from sphericity \cite{Nozawa:1990gt,Buchmann:2002xq,Pascalutsa:2006up}. We 
predict a prolate (positive $G_{E_2}$) shape for the charge distribution in 
$\Sigma^{*~+}$ and oblate (negative sign) for the $\Sigma^{*~-}$. In the case of 
the $\Sigma^{*~0}$ we even obtain a zero crossing at $Q^2\sim 
700$MeV, the charge distribution changing from an oblate to a prolate shape as 
the electromagnetic probe increases in energy. It will be interesting to check 
whether this feature survives a more sophisticated calculation in an extension 
of the present framework including meson cloud effects. Furthermore, a remark on 
the calculation of $G_{E_2}$ for small values of $Q^2$ is in order: it can be seen from 
Eq.~\Eq{eq:expressionGs2}, that the equation to extract $G_{E_2}$ from traces of 
the current \eqref{eq:RS_current} features a $Q^{-4}$ factor which should be 
compensated by cancellations of other terms in order to give a finite value of 
$G_{E_2}$ at $Q^2\sim0$. With the accuracy of our current calculations this 
cancellation is not exact and numerical noise appears. This is manifested in the 
plots by the coloured bands stopping at a small but non-zero value of $Q^2$ (see 
also \cite{Sanchis-Alepuz:2013iia}).

A similar discussion applies for the electric monopole and quadrupole form factors 
of the $\Xi^{*~0}$ $\Xi^{*~-}$ hyperons from Fig.~\ref{fig:xi32_GE}. The form factors 
of $\Xi^{*~0}$ are not vanishing and $G_{E_0}$ is of the same order as $G_E$ for $\Xi^0$. 
The quadrupole form factor $G_{E_2}$ also features a zero crossing for the $\Xi^{*~0}$, 
but this time in the region in which the singularities of the quark propagator are 
probed (see discussion above and Appendix~\ref{sec:kinematics}) and our 
predictions are less reliable.

From the electric monopole form factors we can extract the electric radii. Our results are given in Tab.~\ref{tab:radii_decuplet}. As already mentioned in the previous section, all strange members of the decuplet have a similar or slightly smaller size that their octet counterparts. It will be interesting to study whether the different effect of the meson cloud in octet and decuplet states alters this feature.
\begin{table}[t!]
\caption{Electric-$E_0$ and magnetic-$M_1$ mean-square radii of the decuplet hyperons 
in units of (fm)$^2$.}
\label{tab:radii_decuplet}       
\begin{tabular}{cccccc}
\hline\noalign{\smallskip}
~ & $\Sigma^{*+}$ & $\Sigma^{*0}$ & $\Sigma^{*-}$ & $\Xi^{*0}$ & $\Xi^{*-}$ \\
\noalign{\smallskip}\hline\noalign{\smallskip}
$\langle r^2_{E0} \rangle$ & 0.55(4) & 0.04(1) & 0.51(5) & 0.08(6) & 0.38(7)  \\
$\langle r^2_{M1} \rangle$ & 0.42(5) & 1.36(4) & 0.31(4) & 0.38(7) & 0.41(10)  \\
\noalign{\smallskip}\hline
\end{tabular}
\end{table}

The magnetic dipole $G_{M_1}$ and octupole $G_{M_3}$ form factors are shown in the two lower panels of 
Figs.~\ref{fig:sigma32_GE} and \ref{fig:xi32_GE} for the $\Sigma^*$ triplet and the $\Xi^*$ doublet 
respectively. The most interesting feature is again a zero crossing of the dipole form factor for 
the neutral members $\Sigma^{*~0}$ and $\Xi^{*~0}$. 

From the magnetic dipole $G_{M_1}$ form factor we extract the magnetic radii, shown 
in Tab.~\ref{tab:radii_decuplet}. The most remarkable and puzzling feature of this 
calculation is the enormous magnetic radius of the neutral $\Sigma^{*~0}$. 
The reason for this deviation in comparison with the other decuplet states is 
not clear to us. Note that also in a quark model calculation of these 
observables 
\cite{Wagner:2000ii}, the $\Sigma^{*~0}$ (together with the $\Xi^{*~0}$) has the 
largest magnetic radius of all ground-state hyperons. The differences are, however, 
much smaller than in our calculation.

Finally, with respect to the magnetic octupole form factors, the factor $Q^{-6}$ 
in the extraction of the form factor \Eq{eq:expressionGs4} prevents us, with our 
current accuracy, to extract precise results. The only feature that seems robust 
at the present stage is the sign of the corresponding form factor. As with 
$G_{E_2}$, a non-vanishing $G_{M_3}$ indicates a deformation of the magnetic 
distribution, with a positive (negative) sign signalling a prolate (oblate) 
shape. 

\section{Summary}\label{sec:summary}

We have presented results for the elastic electromagnetic form factors of all ground-state hyperons 
in the rainbow ladder truncation of the covariant three-body Bethe-Salpeter equation. Our results 
have been compared with lattice QCD data where available, showing good agreement with small
deviations in some isospin channels that can be explained. At vanishing momentum transfer, where 
experimental data exists for the members of the isospin octet, is where the largest deviations appear. 
We interpret this as a manifestation of the absence of a pion and kaon cloud in the present calculation. 
Similar discrepancies have been observed previously in the calculation of non-strange baryon form 
factors \cite{Eichmann:2011vu,Eichmann:2011pv,Sanchis-Alepuz:2013iia}.
Including such effects is a necessary next step and work in this direction based 
on \cite{Fischer:2008sp,Fischer:2008wy,Sanchis-Alepuz:2014wea} is in progress. 

\section{Acknowledgements}
This work has been supported by an Erwin Schr\"odinger fellowship J3392-N20 
from the Austrian Science Fund, FWF, by the Helmholtz International Center 
for FAIR within the LOEWE program of the State of Hesse, and by the 
DFG collaborative research centre TR 16.

\appendix

\section{Fits for form factors}

In order to facilitate the use of our results for other calculations, we have 
fitted them using the following general rational form 
\begin{equation}\label{eq:fits}
 G(Q^2)=\frac{n_0+n_1Q^2}{1+d_1Q^2+d_2Q^4+d_3Q^6+d_4Q^8}~.
\end{equation}
Our aim has been to use the polynomials of lowest order possible in the 
denominator that reproduce the calculated form factors. The resulting fitting 
parameters are shown in tables \ref{tab:fit_octet_elastic} and 
\ref{tab:fit_decuplet_elastic}.

\begin{table*}
\begin{center}
\begin{tabular}{c||ll|ll}
 & $G_E~(\eta=1.6)$ & $G_E~(\eta=2.0)$ & $G_M~(\eta=1.6)$ & $G_M~(\eta=2.0)$ \\
\hline
\hline
\multirow{3}{*}{$\Sigma^+$} & $n_0=1$ & $n_0=1$ & $n_0=2.139$ & $n_0=2.082$ \\
                            & $d_1=2.249$ & $d_1=2.503$ & $d_1=1.767$ & $d_1=1.944$ \\
                            & $d_2=1.528$ & $d_2=1.981$ & $d_2=0.654$ & $d_2=0.607$ \\                            
\hline
\multirow{5}{*}{$\Sigma^0$} & $n_1=-0.209$ & $n_1=-0.284$ & $n_0=0.606$ & $n_0=0.603$ \\
                            & $d_1=1.766$ & $d_1=2.914$ & $d_1=1.575$ & $d_1=1.792$ \\
                            & $d_2=2.666$ & $d_2=2.351$ & $d_2=0.608$ & $d_2=0.555$ \\                            
                            & $d_3=-0.872$ & $d_3=-0.327$ &           &             \\                            
                            & $d_4=0.425$ & $d_4=0.282$ &             &             \\                            
\hline
\multirow{3}{*}{$\Sigma^-$} & $n_0=-1$ & $n_0=-1$ & $n_0=-0.928$ & $n_0=-0.877$ \\
                            & $d_1=1.801$ & $d_1=2.020$ & $d_1=2.044$ & $d_1=2.166$ \\
                            & $d_2=0.651$ & $d_2=0.641$ & $d_2=0.743$ & $d_2=0.710$ \\                            
\hline
\multirow{3}{*}{$\Xi^0$}    & $n_1=-0.473$ & $n_1=-0.434$ & $n_0=-1.383$ & $n_0=-1.399$ \\
                            & $d_1=2.886$ & $d_1=0.513$ & $d_1=1.363$ & $d_1=1.582$ \\
                            & $d_2=2.253$ & $d_2=3.135$ & $d_2=0.355$ & $d_2=0.346$ \\                            
\hline
\multirow{3}{*}{$\Xi^-$}    & $n_0=-1$ & $n_0=-1$ & $n_0=-0.705$ & $n_0=-0.613$ \\
                            & $d_1=1.450$ & $d_1=1.774$ & $d_1=0.928$ & $d_1=0.783$ \\
                            & $d_2=0.634$ & $d_2=0.588$ & $d_2=0.168$ & $d_2=0.172$ \\  
\hline
\multirow{3}{*}{$\Lambda$}  & $n_1=-0.092$ & $n_1=-0.203$ & $n_0=-0.505$ & $n_0=-0.498$ \\
                            & $d_1=-1.732$ & $d_1=0.049$ & $d_1=0.826$ & $d_1=0.971$ \\
                            & $d_2=6.262$ & $d_2=4.056$ & $d_2=0.757$ & $d_2=0.339$ \\                             
\hline
\end{tabular}
\caption{Fitting parameters for the elastic form factors of the strange members of the baryon octet.}\label{tab:fit_octet_elastic}
\end{center}
\end{table*}

\begin{table*}
\begin{center}
\begin{tabular}{c||ll|ll|ll}
 & $G_{E0}~(\eta=1.6)$ & $G_{E0}~(\eta=2.0)$ & $G_{E2}~(\eta=1.6)$ & $G_{E2}~(\eta=2.0)$ & $G_{M1}~(\eta=1.6)$ & $G_{M1}~(\eta=2.0)$ \\
\hline
\hline
\multirow{4}{*}{$\Sigma^{*+}$} & $n_0=1$     & $n_0=1$     & $n_0=-1.040$ & $n_0=-0.393$ & $n_0=3.392$  & $n_0=3.606$     \\
                               & $d_1=2.197$ & $d_1=2.514$ & $n_1=-0.555$ & $n_1=-0.248$ & $d_1=2.766$  & $d_1=3.590$     \\
                               & $d_2=2.897$ & $d_2=5.875$ & $d_1=1.932$  & $d_1=0.123$  & $d_2=0.940$  & $d_2=2.011$     \\
                               &             &             & $d_2=3.765$  & $d_2=3.812$  &              &    \\                            
\hline
\multirow{6}{*}{$\Sigma^{*0}$} & $n_1=-0.116$ & $n_1=-0.208$ & $n_0=-0.097$ & $n_0=-0.066$ &  $n_0=0.190$  & $n_0=0.250$  \\
                               & $d_1=-1.341$ & $d_1=-0.872$ & $n_1=0.172$  & $n_1=0.079$  &  $n_1=-0.701$ & $n_1=-0.851$ \\
                               & $d_2=3.985$  & $d_2=4.498$  & $d_1=-1.637$ & $d_1=-0.159$ &  $d_1=4.020$  & $d_1=2.316$  \\                            
                               &              &              & $d_2=10.525$ & $d_2=4.418$  &  $d_2=0.031$  & $d_2=3.665$  \\                            
                               &              &              &              &              &  $d_3=3.550$  & $d_3=0.600$  \\                            
                               &              &              &              &              &  $d_4=-0.567$ & $d_4=-0.020$ \\ 
\hline
\multirow{4}{*}{$\Sigma^{*-}$} & $n_0=-1$     & $n_0=-1$     & $n_0=0.762$  & $n_0=0.259$  & $n_0=-2.929$  & $n_0=-2.965$ \\
                               & $d_1=1.938$  & $d_1=2.402$  & $n_1=0.312$  & $n_1=0.362$  & $d_1=1.939$   & $d_1=2.642$  \\
                               & $d_2=1.326$  & $d_2=1.434$  & $d_1=1.041$  & $d_1=0.028$  & $d_2=0.539$   & $d_2=0.579$  \\
                               &              &              & $d_2=2.160$  & $d_2=3.580$  &               &              \\                       
\hline
\multirow{4}{*}{$\Xi^{*0}$}    & $n_1=-0.114$ & $n_1=-0.608$ & $n_0=-0.185$ & $n_0=-0.109$ &  $n_0=0.300$  & $n_0=0.421$  \\
                               & $d_1=-1.190$ & $d_1=0.769$  & $n_1=0.273$  & $n_1=0.170$  &  $n_1=-0.658$ & $n_1=-1.347$ \\
                               & $d_2=1.599$  & $d_2=4.455$  & $d_1=-0.374$ & $d_1=-0.409$ &  $d_1=-0.506$ & $d_1=0.868$  \\                            
                               &              &              & $d_2=3.780$  & $d_2=3.484$  &  $d_2=2.447$  & $d_2=3.621$  \\                            
\hline
\multirow{4}{*}{$\Xi^{*-}$}    & $n_0=-1$     & $n_0=-1$     & $n_0=0.604$  & $n_0=0.242$  & $n_0=-2.867$  & $n_0=-2.878$ \\
                               & $d_1=1.336$  & $d_1=1.927$  & $n_1=0.360$  & $n_1=0.292$  & $d_1=1.315$   & $d_1=2.168$  \\
                               & $d_2=0.989$  & $d_2=0.910$  & $d_1=0.364$  & $d_1=0.317$  & $d_2=0.462$   & $d_2=0.343$  \\
                               &              &              & $d_2=1.916$  & $d_2=2.301$  &               &              \\                       
\hline
\end{tabular}
\caption{Fitting parameters for the elastic form factors of the strange members of the baryon decuplet.}\label{tab:fit_decuplet_elastic}
\end{center}
\end{table*}

\section{Extraction of form factors}

Given the current corresponding to a spin-$\nicefrac{1}{2}$ baryon, 
Eq.~\Eq{eq:D_current}, the form factors are extracted using the following 
expressions
\begin{align}
G_{E} &= \frac{1}{2i\sqrt{1+\tau}}\textrm{Tr}\{J^\mu\hat{P}^\mu\}\,, \label{eq:expressionGE}\\
G_{M} &= \frac{i}{4\tau}\textrm{Tr}\{J^\mu\gamma_T^\mu\}\,, \label{eq:expressionGM}
\end{align} 
with $\gamma_T$ the transverse projection of the Dirac matrices with respect to $P$ and $\tau=Q^2/4M^2$.

In the case of spin-$\nicefrac{3}{2}$ baryons, given the current 
\Eq{eq:RS_current}, the form factors are extracted via
\begin{align}
G_{E_0} &= \frac{s_2-2s_1}{4i\sqrt{1+\tau}}\,, \label{eq:expressionGs1}\\
G_{M_1} &= \frac{9i}{40\,\tau}\left(s_4-2s_3\right)\,, \label{eq:expressionGs2}\\
G_{E_2} &= \frac{3}{8i\,\tau^2\sqrt{1+\tau}} \left[ 2s_1
\left(\tau+\frac{3}{2}\right) - \tau s_2 \right], \label{eq:expressionGs3}\\
G_{M_3} &= \frac{3i}{16\,\tau^3} \left[ 2s_3 \left(\tau+\frac{5}{4}\right) -
\tau s_4 \right]~,\label{eq:expressionGs4}
\end{align} 
where the traces $s_i$  are
\begin{align}
s_1 (\tau) &= \textnormal{Tr}\left\{ J^{\mu,\alpha\beta}  \hat{P}^\mu  \hat{P}^\alpha 
\hat{P}^\beta \right\}~, \label{sses1}\\
s_2 (\tau) &= \textnormal{Tr}\left\{ J^{\mu,\alpha\alpha}  \hat{P}^\mu \right\}~, \label{sses2}\\
s_3 (\tau) &= \textnormal{Tr}\left\{ J^{\mu,\alpha\beta} \,\gamma^\mu_T  \hat{P}^\alpha
 \hat{P}^\beta \right\}~, \label{sses3}\\
s_4 (\tau) &= \textnormal{Tr}\left\{ J^{\mu,\alpha\alpha} \,\gamma^\mu_T \right\} \,\label{sses4}.
\end{align} 

\section{Flavour traces for form factors}\label{sec:flavour}
\begin{table*}[t!]
 \begin{center}
 \small
\renewcommand{\arraystretch}{1.2}
  \begin{tabular}{lcc}\hline
 state 		& A  										&  S   															\\ \hline\hline
$p$  		& $\frac{1}{\sqrt{2}}\left(udu-duu\right)$  & $\frac{1}{\sqrt{6}}\left(2uud-udu-duu\right)$ 				\\[1.0ex]
$n$  		& $\frac{1}{\sqrt{2}}\left(udd-dud\right)$  & $\frac{1}{\sqrt{6}}\left(udd+dud-2ddu\right)$					\\[1.0ex]
$\Sigma^+$ 	& $\frac{1}{\sqrt{2}}\left(usu-suu\right)$ 	& $\frac{1}{\sqrt{6}}\left(2uus-usu-suu\right)$					\\[1.0ex]
$\Sigma^0$ 	& $\frac{1}{2}\left(usd+dsu-sud-sdu\right)$ & $\frac{1}{\sqrt{12}}\left(2uds+2dus-usd-dsu-sud-sdu\right)$ 	\\[1.0ex]
$\Sigma^-$ 	& $\frac{1}{\sqrt{2}}\left(dsd-sdd\right)$ 	& $\frac{1}{\sqrt{6}}\left(2dds-dsd-sdd\right)$ 				\\[1.0ex]
$\Xi^0$ 	& $\frac{1}{\sqrt{2}}\left(uss-sus\right)$ 	& $\frac{1}{\sqrt{6}}\left(uss+sus-2ssu\right)$ 				\\[1.0ex]
$\Xi^-$ 	& $\frac{1}{\sqrt{2}}\left(dss-sds\right)$ 	& $\frac{1}{\sqrt{6}}\left(dss+sds-2ssd\right)$ 				\\ [1.0ex]
$\Lambda^0$ & $\frac{1}{\sqrt{12}}\left(2uds-2dus+sdu-dsu+usd-sud\right)$ & $\frac{1}{2}\left(usd+sud-dsu-sdu\right)$ 	\\[1.0ex]
\hline
\end{tabular}
\caption{Baryon octet flavour amplitudes.\vspace*{5mm}
\label{tab:octet_flavour}}
 \end{center}
\end{table*}

For the calculation of form factors we need the charge matrices $\mathcal{Q}$, 
Eq.~\Eq{eq:charge_matrices}. In the case of Eq.~\Eq{eq:FFeqRL_simple} we also need the matrices
$\mathcal{F}_1$ and $\mathcal{F}_2$, defined as
\begin{align}
 \mathcal{F}^{\rho\rho'}_1=F^\rho_{abc}F^{\rho'}_{bca}~,\\
 \mathcal{F}^{\rho\rho'}_2=F^\rho_{abc}F^{\rho'}_{cab}~.
\end{align}
They depend on the symmetry of the flavour amplitudes only.
For octet baryons, and using the standard flavour amplitudes for 
octet baryons in Tab.~\ref{tab:octet_flavour}, they are given by
\begin{flalign}
 \mathcal{F}_1=\left(
     \begin{array}{cc}
      \nicefrac{-1}{2} & \nicefrac{\sqrt{3}}{2}               \\
       \nicefrac{-\sqrt{3}}{2} & \nicefrac{-1}{2}               
     \end{array}\right)~,
     ~~ \mathcal{F}_2=\left(
     \begin{array}{cc}
      \nicefrac{-1}{2} & \nicefrac{-\sqrt{3}}{2}               \\
       \nicefrac{\sqrt{3}}{2} & \nicefrac{-1}{2}               
     \end{array}\right)~.
\end{flalign}

Finally, denoting with $J^{\rho\rho'}_{\lambda_1\lambda_2\lambda_3}$ the result of the last term in 
\Eq{eq:FFeqRL} for a particular combination of quark flavours 
$(\lambda_1\lambda_2\lambda_3)$, 
\begin{table}[h!]
 \begin{center}
 \small
\renewcommand{\arraystretch}{1.2}
  \begin{tabular}{lc}\hline
 state 			& S    											\\ \hline\hline
$\Delta^{++}$  	& $uuu$ 										\\[0.7ex]
$\Delta^{+}$  	& $\frac{1}{\sqrt{3}}\left(uud+udu+duu\right)$ 	\\[0.7ex]
$\Delta^{0}$ 	& $\frac{1}{\sqrt{3}}\left(udd+dud+ddu\right)$ 	\\[0.7ex]
$\Delta^{-}$ 	& $ddd$ 										\\[0.7ex]
$\Sigma^{*+}$ 	& $\frac{1}{\sqrt{3}}\left(uus+usu+suu\right)$ 	\\[0.7ex]
$\Sigma^{*0}$ 	& $\frac{1}{\sqrt{6}}\left(uds+usd+dus+dsu+sud+sdu\right)$ \\[0.7ex]
$\Sigma^{*-}$ 	& $\frac{1}{\sqrt{3}}\left(dds+dsd+sdd\right)$ 	\\ [0.7ex]
$\Xi^{*0}$ 		& $\frac{1}{\sqrt{3}}\left(uss+sus+ssu\right)$ 	\\[0.7ex]
$\Xi^{*-}$ 		& $\frac{1}{\sqrt{3}}\left(dss+sds+ssd\right)$ 	\\[0.7ex]
$\Omega^{-}$ 	& $sss$ \\
\hline
\end{tabular}
\caption{Baryon decuplet flavour amplitudes.
\label{tab:decuplet_flavour}}
 \end{center}
\end{table}
we get as a result for 
Eq.~\Eq{eq:FFeqRL_simple} (omitting all other indices for clarity)
\begin{align}
&J_p=2J^{11}_{uuu}~,\\
&J_n=-J^{11}_{uuu}+J^{22}_{uuu}~,\\
&J_{\Sigma^+}=J^{11}_{suu}+J^{11}_{usu}+\frac{1}{3} \left(-\sqrt{3} 
J^{12}_{suu}+\sqrt{3}J^{12}_{usu}-\sqrt{3}J^{21}_{suu}+\sqrt{3} 
J^{21}_{usu}+J^{22}_{suu}+J^{22}_{usu}-2J^{22}_{uus}\right)~,\\
&J_{\Sigma^0}=\frac{1}{12} \left(3 J^{11}_{suu}+3 J^{11}_{usu}-\sqrt{3} 
J^{12}_{suu}+\sqrt{3} J^{12}_{usu}-\sqrt{3} J^{21}_{suu}+\sqrt{3} 
J^{21}_{usu}+J^{22}_{suu}+J^{22}_{usu}-8J^{22}_{uus}\right)~,\\
&J_{\Sigma^-}=\frac{1}{6} \left(-3 J^{11}_{suu}-3 J^{11}_{usu}+\sqrt{3} 
J^{12}_{suu}-\sqrt{3} J^{12}_{usu}+\sqrt{3} J^{21}_{suu}-\sqrt{3} 
J^{21}_{usu}-J^{22}_{suu}-J^{22}_{usu}-4 J^{22}_{uus}\right)~,\\
&J_{\Xi^0}=\frac{1}{6} \left(-3 J^{11}_{sus}-3 J^{11}_{uss}-\sqrt{3} 
J^{12}_{sus}+\sqrt{3} J^{12}_{uss}-\sqrt{3} J^{21}_{sus}+\sqrt{3} 
J^{21}_{uss}+8 J^{22}_{ssu}-J^{22}_{sus}-J^{22}_{uss}\right)~,\\
&J_{\Xi^-}=\frac{1}{6} \left(-3J^{11}_{sus}-3 J^{11}_{uss}-\sqrt{3} 
J^{12}_{sus}+\sqrt{3} J^{12}_{uss}-\sqrt{3} J^{21}_{sus}+\sqrt{3} 
J^{21}_{uss}-4J^{22}_{ssu}-J^{22}_{sus}-J^{22}_{uss}\right)~,\\
&J_{\Lambda}=\frac{1}{12} \left(J^{11}_{suu}+J^{11}_{usu}-8 
J^{11}_{uus}+\sqrt{3} J^{12}_{suu}-\sqrt{3} J^{12}_{usu}+\sqrt{3} 
J^{21}_{suu}-\sqrt{3}
J^{21}_{usu}+3 (J^{22}_{suu}+J^{22}_{usu})\right)~.
\end{align}

In the case of decuplet baryons, $\mathcal{F}_1$ and $\mathcal{F}_2$ are the unit $1\times 1$ matrix and
the combination of charge matrices, using the flavour amplitudes in Tab.~\ref{tab:decuplet_flavour}, give
\begin{align}
&J_{\Delta}=Q_\Delta J_{uuu}~,\\
&J_{\Sigma^{*+}}=\frac{2 (J_{suu}+J_{usu})}{3}-\frac{J_{uus}}{3}~,\\
&J_{\Sigma^{*0}}=\frac{1}{6} (J_{suu}+J_{usu}-2 J_{uus})~,\\
&J_{\Sigma^{*-}}=\frac{1}{3} (-J_{suu}-J_{usu}-J_{uus})~,\\
&J_{\Xi^{0+}}=\frac{1}{3} (2 J_{ssu}-J_{sus}-J_{uss})~,\\
&J_{\Xi^{*-}}=\frac{1}{3} (-J_{ssu}-J_{sus}-J_{uss})\,.
\end{align}

\section{Kinematics}
\label{sec:kinematics}

The relative $p,q$ and total $P$ momenta in the Faddeev amplitude 
\Eq{eq:BSE_amplitude} are defined in terms of
the three quark momenta $p_1$, $p_2$ and $p_3$ as
\begin{align}\label{eq:defpq}
        p &= (1-\zeta)\,p_3 - \zeta (p_1+p_2)\,, &  p_1 &=  -q -\dfrac{p}{2} +
\dfrac{1-\zeta}{2} P~, \nonumber\\
        q &= \dfrac{p_2-p_1}{2}\,,         &  p_2 &=   q -\dfrac{p}{2} +
\dfrac{1-\zeta}{2} P~,\nonumber\\
        P &= p_1+p_2+p_3\,,                &  p_3 &=   p + \zeta  
P~,
\end{align}
with $\zeta=1/3$ a momentum partitioning parameter. The internal quark 
propagators in the Faddeev equation \Eq{eq:Faddeev_coeff3} depend on 
the internal quark momenta
$k_i=p_i-k$ and $\tilde{k}_i=p_i+k$, with $k$ the exchanged momentum. The 
internal
relative momenta, for each of the three terms in the Faddeev equation, are
\begin{align}\label{internal-relative-momenta}
p^{(1)} &= p+k,& p^{(2)} &= p-k,& p^{(3)} &= p,\nonumber\\
q^{(1)} &= q-k/2,& q^{(2)} &= q-k/2, & q^{(3)} &= q+k\,\,.\nonumber\\
\end{align}

Exploiting the symmetries of the problem, the Faddeev equation can be written in 
terms of one diagram only, see Eq.~\Eq{eq:Faddeev_coeff}. The other two terms 
are
obtained from the former by evaluating it at the permuted momenta 
\begin{flalign}
p'=-q-\frac{p}{2}~,\quad q'=-\frac{q}{2}+\frac{3p}{4}~,\nonumber \\
p''=q-\frac{p}{2}~,\quad q''=-\frac{q}{2}-\frac{3p}{4}~,
\end{flalign}
and rotating it with the matrices
\begin{flalign}\label{eq:rotation_matrices} 
H_1^{ij}&=\left[\bar{\tau}^i_{\beta\alpha\mathcal{I}\gamma}(p,q,P)\tau^j_{
\beta\gamma\alpha\mathcal{I}}(p',q',P)\right]~,\\
H_2^{ij}&=\left[\bar{\tau}^i_{\beta\alpha\mathcal{I}\gamma}(p,q,P)\tau^j_{
\gamma\alpha\beta\mathcal{I}}(p'',q'',P)\right]~.
\end{flalign}

Each term carries an associated flavour matrix
\begin{flalign}\label{eq:flavour_matrices}
\mathcal{F}_1^{\rho\rho',\lambda}=&F^{\rho}_{abc}k^F_{bb'cc'}F^{\rho',
\lambda}_{b'c'a}~,\nonumber\\
\mathcal{F}_2^{\rho\rho',\lambda}=&F^{\rho}_{abc}k^F_{aa'cc'}F^{\rho',
\lambda}_{c'a'b}~,\\
\mathcal{F}_3^{\rho\rho',\lambda}=&F^{\rho}_{abc}k^F_{aa'bb'}F^{\rho',
\lambda}_{a'b'c}~.\nonumber
\end{flalign}
with $k_F$ the flavour part of the interaction kernel which, in case of the RL 
kernel is just the unit matrix and the flavour amplitudes $F$ for each baryon 
are given in Tables \ref{tab:octet_flavour} and \ref{tab:decuplet_flavour}.

In the covariant BSE formalism one usually works in Euclidean space-time. The 
on-shell condition $P^2=-M^2$, for $M$ the bound-state mass, thus implies that 
$P$ must be complex. For example, in the bound-state's rest frame we write 
$P=(0,0,0,iM)$ which, from \Eq{eq:defpq}, implies for example
\begin{align}\label{eq:squared_propagator_momentum}
    p_3^2=p^2-\zeta^2M^2+2i\zeta M\sqrt{p^2}\hat{P}\cdot\hat{p}~.
\end{align}
That is, the momenta of the quarks become complex and the quark propagator must 
be known in a region of the complex plane.
Quark propagators typically show complex conjugate poles in the left-half of 
the complex plane (see, e.g., \cite{Dorkin:2013rsa}). From 
\Eq{eq:squared_propagator_momentum} it is easy to see 
that above a certain bound-state mass one encounters such poles.

In the calculation of form factors we work in the Breit frame; that is, the 
photon momentum is defined as \linebreak $Q=(0,0,Q,0)$, or $Q=Q\widehat{e_3}$. The 
incoming and outgoing baryon momenta are then $P_{i/f}=P\mp Q/2$, with
\begin{align}
P=iM\sqrt{1+\tau}\widehat{e_4}~,
\end{align}
and
\begin{align}
\tau=\frac{Q^2}{4M^2}~.
\end{align}
The $\kappa$-th quark momenta before and after the photon coupling are 
$p_\kappa^{i/f}=p_k\mp Q/2$. Now, if we take as an example $p_3=p+\zeta P$, 
then $p_3^{i/f}$ squared can be written as
\begin{align}\label{eq:squared_propagator_momentum_FF}
    (p_3^{i/f})^2=\left(t+iM\zeta \sqrt{1+\tau}\right)~,
\end{align}
with $t$ real. Clearly, for a fixed mass $M$, it is now the photon momentum $Q$ 
via $\tau$ that controls whether or not the poles of quark propagator are 
probed in the calculation of form factors.

%
%

\end{document}